\documentclass[12pt,a4paper]{article}
\usepackage[toc,page]{appendix}
\usepackage{fullpage}
\usepackage[latin1]{inputenc}
\usepackage[french,english] {babel}
\usepackage{amsmath}
\usepackage{amsfonts}
\usepackage{amssymb}
\usepackage{graphicx}
\usepackage{epstopdf}
\usepackage{color}
\usepackage{listings}
\usepackage[stable]{footmisc}
\usepackage{verbatim}
\usepackage{subfig}
\usepackage[top=2.5cm,bottom=2.5cm,right=2.5cm,left=2.5cm]{geometry}
\usepackage{amsthm}
\usepackage{feynmp}
\usepackage{array}
\usepackage{hyperref}
\definecolor{vert}{rgb}{0,0.5,0}
\newcommand{\beqn}{\begin{eqnarray}}
\newcommand{\eeqn}{\end{eqnarray}}
\newcommand{\beqas}{\begin{eqnarray*}}
\newcommand{\eeqas}{\end{eqnarray*}}
\newcommand{\beq}{\begin{equation}}
\newcommand{\eeq}{\end{equation}}
\newcommand{\bit}{\begin{itemize}}
\newcommand{\eit}{\end{itemize}}
\newcommand{\bseq}{\begin{subequations}}
\newcommand{\bal}{\begin{align}}
\newcommand{\eal}{\end{align}}
\newcommand{\eseq}{\end{subequations}}
\newcommand{\nn}{\nonumber}

\begin{document}
\begin{flushright}
\end{flushright}
\vskip 1cm

\begin{center}
{\huge \bf Exploring variations in the gauge sector of a six-dimensional
flavour model }
\vskip .7cm
{\large Jean-Marie Fr\`ere${}^a$}, {\large
Maxim Libanov${}^{b,c}$}, {\large Simon Mollet${}^a$} and {\large Sergey
Troitsky${}^b$}
\vskip .7cm
\emph{${}^a$Service de Physique Th\'eorique,\\
Universit\'e Libre de Bruxelles, ULB, 1050 Brussels, Belgium
\vskip 0.2cm
${}^b$Institute for Nuclear Research of the Russian Academy of Sciences,\\
60th October Anniversary Prospect 7a, 117312, Moscow, Russia
\vskip 0.2cm
${}^{c}$  Moscow Institute of Physics and Technology,\\
Institutskii per., 9, 141700, Dolgoprudny, Moscow Region, Russia}
\end{center}

\begin{abstract}
We address the question, how general is the gauge sector in
extra-dimensional models which explain hierarchies of masses and mixings
of quarks, charged leptons and neutrinos in terms of a single
family of multidimensional fermions. We give qualitative arguments that
though there are a plethora of possible variations, they do not result in
drastical changes of phenomenology.
\end{abstract}

\bigskip

\section{Introduction}
The wonderful world of large and infinite extra dimensions (ED), where
low-energy excitations of multidimensional fields (``zero modes'') are
bound to a (3+1)-dimensional manyfold (``the brane'') representing our
world, was discovered for theoretical physicists in independent works of
Rubakov and Shaposhnikov \cite{RuSha1}, Akama \cite{Akama} and Visser
\cite{Visser} more than four decades ago. Since then, enlarged symmetries
of multidimensional worlds have been exploited in field-theory frameworks
to address various fine-tuning and hierarchy problems of the Standard
Model (SM) of particle physics, see e.g.\ reviews \cite{Ru-UFN, 0907.3074}
and references therein. One of the approaches transfers
geometric symmetries of the extra dimensions into flavour symmetries of
our world, explaining in an elegant way the hierarchy of masses and
mixings of SM quarks and charged leptons~\cite{Libanov:01, Frere:01,
Frere:03} and leading to rich testable phenomenology~\cite{Libanov:02,
Frere:04, Frere:04b, Libanov:05}. The same model explains as well a very
different pattern of neutrino masses and mixing, the difference with
quarks being caused by the Majorana form of the neutrino mass term
\cite{Frere:10} (see Ref.~\cite{Frere:13} for a recent update). The
purpose of the present work is to explore some ways beyond the simplest
model and to sketch how robust its predictions are.

In ED models that hope to embed the SM, some vector fields must be
introduced which will play the role of usual gauge fields at low energy.
Their (almost) massless ``zero'' modes
appear as the usual (3+1)-dimensional (4D) gauge bosons. The way of
implementing a mechanism responsible for that is not always an easy task
for there are further requirements to build a realistic model. Indeed,
while we want the gauge zero mode to interact properly with the fermionic
ones, we know that there will also exist a set of heavier (excited) modes
which should not talk \textit{too much} with this low energy sector,
\textit{i.e.} either there must exist a mass gap or these modes must only
interact very weakly with the low-energy sector \cite{Daemi:03}. On the
other hand, these new modes could manifest themselves at higher energy (in
collider experiments for instance) or in (very) rare processes (e.g.,
flavour-changing neutral currents), thus providing hints for this kind of
models.

In this note, we would like to provide with a short update of the
constraints from these experiments for various models of
this kind. We will focus
on a particular class in (4+2) dimensions where a Nielsen-Olesen vortex-like
defect plays the role of our 4D world \cite{Libanov:01, Frere:01,
Frere:03, Daemi:03, Giovannini:01}. We know that, quite generally in this
background, we can get several localized (chiral) fermionic zero modes from
a single spinor in 6D \cite{Jackiw:81}, each of them associated with a
different winding in ED\footnote{The exact values of the windings are not
important. What will be really relevant for us are the difference in
windings between two modes.}
($e^{iw\varphi},e^{i(w+1)\varphi},e^{i(w+2)\varphi,...}$). They can
acquire (small) masses through the vacuum expectation value of a
Brout-Englert-Higgs (BEH) field $H$. In a certain range of parameters
\cite{Libanov:05}, the particular shape of this vev in ED (non zero in the
core, almost zero outside) leads to a hierarchical pattern of masses. This
idea was exploited in different contexts to reproduce the three SM
generations and their spectrum.
Here
however we will only be interested in their interactions with gauge bosons
(both zero and heavy modes).

In section \ref{sec:Examples}, we come back on some possible ways of
introducing gauge bosons in the model and try to convince the reader that
the expected phenomenology should not change drastically from one
realisation to the other. In particular we will recall the existence of
heavy localized modes whose mass scale is set by the geometry. Unlike the
zero mode, the former possess non zero windings and can therefore be
responsible for flavour changing processes (even in the absence of mixing
in the fermionic sector) \cite{Frere:04, Frere:04b}. In section
\ref{sec:FlavProc}, we comment on these processes and provide with some
numerical results for the precise realization of \cite{Frere:13}. Finally
we conclude in section \ref{sec:Conc}.

\section{Some generic examples}\label{sec:Examples}

Let us here quickly remind some general results. We will focus on models
with 4D Poincar\'e invariance and 4D flat space. The most general metrics
of such kind can be written as \cite{Daemi:02}:
\beq
\mathrm{d}s^2=G_{AB}\mathrm{d}x^A\mathrm{d}x^B=\sigma(y)\eta_{\mu\nu}\mathrm{d}x^{\mu}\mathrm{d}x^{\nu}-\gamma_{ab}(y)\mathrm{d}y^a\mathrm{d}y^b\label{GenMet}
\eeq
With the following choice of gauge:
\beqn
\left\{\begin{aligned}
& \partial_0 W_0-\partial_i W_i=0,\\
& \frac{\partial_a\left(\sqrt{\vert
G\vert}\sigma^{-1}\gamma^{ab}W_b\right)}{\sqrt{\vert
G\vert}\sigma^{-1}}=0,\\
\end{aligned}
\right.\nn
\eeqn
we have the obvious separation of variables in the equation of motion (EOM) for vector modes:
\beq
W_{\mu}(x,y)=\sum_n \omega_{\mu;n}(x)P_n(y)\nn
\eeq
with the modal wavefunctions $P_n$ satisfying
\beq
\frac{\partial_a\left(\sqrt{\vert G\vert}\sigma^{-1}\gamma^{ab}\partial_b P\right)}{\sqrt{\vert G\vert}\sigma^{-1}}
+\sigma^{-1}m^2 P =0\nn.
\eeq
There always exists a zero mode ($m^2=0$) with a constant transverse
wavefunction ($P(y)=$ const), but we cannot conclude, at this level, if it
is normalizable or not.

Two ways to ensure the normalizability are (i) to
deal with compact ED whose finite volume renders the integral with the
constant delocalized wavefunction bounded, or (ii) to make use of warp
factors
\cite{RuSha2, Gogber, Randall:99, Randall:99b} which will sufficiently
"dilute" the wavefunction, yet yield to a finite
integral\cite{Dubovsky:00, Oda:00}. Note that in the latter case, we can
also consider effective wavefunctions in flat space which include warp
factors and are thus localized from this point of view \cite{Daemi:02}. We
will provide realizations of these two scenarios in the further simplified
metrics, which is a particular case of (\ref{GenMet}):
\beq
\mathrm{d}s^2=\sigma(u)\eta_{\mu\nu}\mathrm{d}x^{\mu}\mathrm{d}x^{\nu}-\mathrm{d}u^2-\gamma(u)\mathrm{d}v^2\nn
.
\eeq

A simple example of the first way (compact space) is the 2-sphere
\cite{Frere:03, Frere:04, Frere:04b} of radius $R$ which corresponds to
$u=R\theta$, $v=R\varphi$ and $\gamma=\sin^2\theta$. The modal equation
becomes then the equation for spherical harmonics with
$R^2m^2=\ell(\ell+1)$. As expected we have a (normalizable) zero mode
$\ell=0$ with constant wavefunction $P=1/\sqrt{4\pi}R$. Heavier modes
appear to be normalizable, too. The mass scale is dictated by the size of
ED. In particular, there is a mass gap of the order $1/R$. For each value
of $\ell$, there are degenerate modes with windings $-\ell \leq m \leq
\ell$. The wavefunctions oscillate on a scale of order $R$ for the
lightest modes.

If we opt instead for the warped case, the warp metrics can be
parametrized \cite{Daemi:03} as $u=r$, $v=a\theta$, $\sigma=e^{A(r)}$ and
$\gamma=e^{B(r)}$. The precise behaviour of the $A$ and $B$ functions are
determined by the exact realization of the defect, but we can establish
general features of their asymptotics by requiring (i) the metrics to be a
regular solution of the 6D Einstein equations where a negative bulk
cosmological constant balances a positive string tension (in the
core)\footnote{Note that at 4D level we ask for a zero cosmological
constant to have a flat space.} and (ii) the gravity to be
localized\footnote{\textit{i.e.} ask for a normalizable zero mode for the
graviton \cite{Giovannini:02}.}. What we get is \cite{Daemi:03,
Giovannini:01} $A'(0)=0$ and $B(r\rightarrow 0)\sim 2 \ln\left(r/a\right)$
around the origin and $A=B=-2rc$ outside the core ($c$ is a dimensional
constant related to the bulk cosmological constant) which correspond to an
AdS${}_6$ geometry. We still have the arbitrariness of normalization and
choose $A(0)=0$. The dimensionfull constant, which will play an important
role later on, $a$ is not a free parameter but is determined by an
interplay between the gravity and the vortex scales. With these asymptotics
it is easy to realize that the two ED are a warped plane in polar
coordinates and it is then obvious to further develop the $P$
wavefunctions on a Fourier basis: \beq P_n(r,\theta)=\sum_{\ell}
\rho_{n\ell}(r)e^{i\ell\theta}\nn.
\eeq
With this, the equation for $\rho$ becomes:
\beq
\rho''+\left(A'+\frac{B'}{2}\right)\rho'+\left(m^2e^{-A}-\frac{\ell^2}{a^2}e^{-B}\right)\rho=0\nn
.
\eeq
Outside the core, the solutions are classified in terms of
$\mu^2=m^2-\ell^2/a^2$. For $\mu=0$, we have a constant solution, while for
$\mu\neq 0$, it reads
\beq
\rho(r)=e^{\frac{3}{2}cr}\left[C_1J_{3/2}\left(\frac{\mu}{c}e^{cr}\right)+C_2Y_{3/2}\left(\frac{\mu}{c}e^{cr}\right)\right]\nn
,
\eeq
where $J$ and $Y$ are Bessel functions, and $C$'s are arbitrary constants.
The boundary conditions (absence of the flux at infinity) lead to a
continuous spectrum for $\mu>0$ \cite{Gherg:00}. If we use the expression
of $J$ and $Y$ in terms of elementary functions, it is easy to show that
$\rho$ behaves as $ \eta e^{cr}$ at sufficiently large $r$, where $\eta$
is some oscillating and bounded function. Now remember that, in the
initial action, we have a factor $\sim\sqrt{\vert
g\vert}(g^{00})^2=ae^{B/2}\sim e^{-cr}$ for the kinetic term of 4D gauge
component (and the integral over $r$ fixes the normalization). As
announced, we can define an effective wavefunction that takes this warp
factor into account, then we can conclude if the associated mode is
localized or not. With the definition $\zeta(r)=e^{-\frac{c}{2}r}\rho(r)$,
we see that for the "constant" mode $\zeta_0(r)\sim e^{-\frac{c}{2}r}$ is
localized\footnote{Note that in the usual 5D Randall-Sundrum models, this
zero mode is not normalizable because the $e^B$ factor is not present. The
presence of an extra warped dimension helps to "dilute" more efficiently
the constant wavefunction.}, while the continuous spectrum $\zeta_c(r)\sim
\eta e^{\frac{c}{2}r}$ is not.
The "not localized" states have most of their weight at large distances
(therefore reducing the overlap). Now near the origin, the regular solution is:
\beq
\rho(r)\sim J_{\ell}(mr),\nn
\eeq
For $m=\ell/a$ (corresponding to localized mode $\mu=0$ at infinity) we have (note that here, the metric factor is simply $r$):
\beq
\rho_0(r)\sim J_{\ell}\left(\ell\frac{r}{a}\right).\nn
\eeq
For $\ell=0$ we get the usual constant solution (which matches with the
constant solution at infinity, since we know that $\rho=$ const is an exact
solution for the all range of $r$). For non-zero $\ell$, we cannot get an
exact solution, but we see that (at least for the first modes) we have
oscillating functions with a scale of oscillation of order $a$.

In conclusion, we have a pattern which looks very much like the spherical
case: discrete (localized) modes with mass scale $1/a$ and this same scale
giving also an idea of the oscillation scale for the associated
wavefunctions. On the other hand, there are (associated to each of these
bounded modes) a continuum, starting just above, but the delocalization
should kill the overlaps with localized profiles. Of course this should be
computed properly to be more quantitative.

\section{Flavour violating processes} \label{sec:FlavProc}

Thanks to the separation of variables, the whole set of modal
wavefunctions can be decomposed as a product of a radial part\footnote{On
the sphere the angle $\theta$ plays the role of the radial variable.} and
an angular one. For the fermion zero modes, the radial part is localized
around the vortex\footnote{Note nevertheless that the size of these
functions must be larger than the size of the vortex in general if we want
to produce a sufficiently strong hierarchy between families (see
\cite{Frere:13} for instance).}, while for the bosonic modes these are
oscillating functions spread in the bulk. In the compactification
procedure (which reduces the complete 6D theory to an effective 4D one
where all modes interact among themselves), the integration over the
radial component controls the strength of the interaction through the
overlaps of wavefunctions, while the one over angular component gives a
selection rule which forbids interactions with non zero total winding
(this can be interpreted as the angular momentum conservation in the ED).

If we neglect mixing between fermions, each family is associated with one and only one winding number $i$. Then the interaction
\beq
\kappa\bar{\psi}_i \gamma^{\mu}\omega_{\mu,m}\psi_{i'}\nn
\eeq
is allowed if and only if $m=i-i'$. The strength $\kappa$ depends on the
radial integral\footnote{In principle, $\kappa$ could be infinitely
reduced by localizing more and more the fermion wavefunctions (through
stronger and stronger interactions with the vortex). However as mentioned
above, we are technically limited because we require (high) hierarchies
between generations. We could still hope to squeeze both fermion and $H$
fields in such a way that the hierarchy is safe, but a detailed analysis
(too technical to be put in here) of the scalar sector (in the spherical
case only, up to now) showed that, once $m_H$ is fixed, we don't have this
freedom anymore. Nevertheless it still is worth looking for smaller
$\kappa$ than imposed by the model, because we do not know what happens in
a different geometry.}. Allowed effective four-fermion interactions,
\beq \frac{\kappa\kappa'}{M^2_{\omega}
}\left(\bar{\psi}_i O \psi_{i'}\right)\left(\bar{\psi}_j
O'\psi_{j'}\right)\nn ,
\eeq correspond to $(i'-i)=(j-j')$, or in other words $\Delta G=0$, if $G$
is some kind of family number. Thus, in first approximation (no mixing),
only $\Delta G=0$ interactions can be observed.

\subsection{Forbidden kaon decays}

The best experimental restriction on flavour violating processes with
$\Delta G=0$ comes from the decay $K_L^0\rightarrow \mu^+e^-$. In SM this
process is suppressed because it is forbidden at the tree level. In our
model however, there is an excited gauge mode which can mediate this
decay.

To be more precise, let us focus on the spherical compactification for
which we have a concrete realisation \cite{Frere:13}.
There, we have presented a set of couplings which reproduce well the SM
masses and mixings as well as satisfy all constraints for masses and
mixings in the neutrino sector, giving some predictions for future
experiments. This realisation of the model has a fixed $R=100$~TeV. Having
all couplings fixed, we can perform quantitative calculations of all
particular processes.

For any neutral gauge field $W_A$ which interacts
with the fermions, we get the following effective Lagrangian at 4D level
(the scalar modes don't interact with SM fermions):
\beq
\mathcal{L}_{4D} \supset \sum_{\ell}\sum_{\underset{\vert n-m\vert \leq \ell}{m,n}} E^{\ell,\vert n-m\vert}_{mn}U^*_{mj}U_{nk}\left(\bar{\psi}_j \gamma^{\mu} Q \psi_k\right) \omega^{(*)}_{\mu;\ell, \vert n-m\vert} \label{eff4D}
\eeq
where $E^{\ell,\vert n-m\vert}_{mn}$ are the results of the overlaps (see
\cite{Frere:04} for details). For $\ell=0$, we have $E^{0,0}_{nn}=1$
(normalization) which permits to identify $Q$ with SM charges. $U$ is the
unitary mixing matrix\footnote{Note that $U$ matrices are not unique.
Indeed, we could as well use $U_L'=U_L\
\text{diag}(e^{i\phi_1},e^{i\phi_2},e^{i\phi_3})$ and $U_R'=U_R\
\text{diag}(e^{i\phi_1},e^{i\phi_2},e^{i\phi_3})$ (with the same phases)
since it doesn't affect the masses, but these are obviously not
physical.}. If it disappears properly for $\ell=0$, this is no more the
case for higher $\ell$'s. Thus, in our model, it makes sense to talk about
mixing in up quarks and down quarks separately, for instance.
$\omega^{(*)}_{\mu}$ are the 4D fields for each modes. When $n-m\neq 0$,
these are complex fields. In our notations, for $n-m>0$ we have to use
$\omega_{\mu}$, so it destroys a mode with winding $\vert n-m\vert$, while
for $n-m<0$ we have to use $\omega^*_{\mu}$, so it creates a mode with
winding $\vert m-n\vert$.

$K_L^0$ is a combination of $\bar{s}d$ and $\bar{d}s$. The first one
corresponds to indices $j=2$ and $k=1$ in (\ref{eff4D}). We can define
matrices $\Omega^{\ell}_{mn}=U^*_{m2}U_{n1}E_{mn}^{\ell,\vert n-m\vert}$
which tell us about the strength of coupling with each mode
$\omega_{\mu;\ell,0}$, $\omega_{\mu;\ell,1}$ and $\omega_{\mu;\ell,2}$.
Note that mixings in left and right sectors are different in general. For
the model of \cite{Frere:13} we have:
\beq
\Omega^{\ell}_L=
\begin{pmatrix}
0.232\ E_{11}^{\ell,0} & -0.057\ E_{12}^{\ell,1} & 0.003\ E_{13}^{\ell,2}\\
0.941\ E_{21}^{\ell,1} & -0.231\ E_{22}^{\ell,0} & 0.013\ E_{23}^{\ell,1}\\
-0.052\ E_{31}^{\ell,2} & 0.013\ E_{32}^{\ell,1} & -0.001\ E_{33}^{\ell,0}
\end{pmatrix}
\quad \text{and}\quad
\Omega^{\ell}_R=
\begin{pmatrix}
0.053\ E_{11}^{\ell,0} & -0.003\ E_{21}^{\ell,1} & 0\\
0.997\ E_{21}^{\ell,1} & -0.053\ E_{22}^{\ell,0} & 0\\
-0.001\ E_{31}^{\ell,2} & 0 & 0
\end{pmatrix}\nn
\eeq
For both matrices, the dominant elements are $(21)$. This means that the
dominant process is the (virtual) creation of a $\omega_{\mu;\ell,1}$ (for
all allowed $\ell$). At first sight, it seems that the contribution to
$\omega_{\mu;\ell,0}$ is significant too. But to be more precise, we have
to evaluate the overlaps $E^{\ell}$ and sum over all contributions. In
particular, the total contribution to $\omega_{\mu;\ell,0}$ is simply the
trace (other can be obtained as sums over elements of lines parallel to
the diagonal). It then is obvious (because of the unitarity of $U$) that
this is negligible as long as $E^{\ell,0}_{11}\simeq
E^{\ell,0}_{22}(\simeq E^{\ell,0}_{33})$. This result is exact for
$\ell=0$ by definition and is expected to be a good approximation for the
first $\ell$'s which correspond to slowly oscillating modes (thus
embracing all fermion wavefunctions in a very similar way). As an example,
we compute the contributions of the first modes in Table
\ref{tab:quarkcoup} (for left-handed quarks only).

\begin{table}[h!]\centering
\begin{tabular}{|c|cccccc|}
\hline\hline
$\ell$            & $0$ & $1$ & $2$ & $3$ &$4$ & $5$ \\ \hline \hline
$E_{11}^{\ell,0}$ & $1$ & $1.004$ & $0.492$ & $0.149$ & $0.014$ & $-0.020$ \\
$E_{22}^{\ell,0}$ & $1$ & $1.073$ & $0.496$ & $0.027$ & $-0.172$ & $-0.206$\\
$E_{33}^{\ell,0}$ & $1$ & $1.419$ & $1.268$ & $0.923$ & $0.603$ & $0.374$\\ \hline
$\omega_{\mu;\ell,0}$ & $0$ & $-0.016$ & $-0.017$ & $0.027$ & $0.042$ & $0.043$ \\ \hline
$E_{12}^{\ell,1}$ & $/$ & $0.780$ & $0.872$ & $0.621$ & $0.359$ & $0.186$ \\
$E_{23}^{\ell,1}$ & $/$ & $0.638$ & $0.908$ & $0.844$ & $0.640$ & $0.440$ \\ \hline
$\omega^*_{\mu;\ell,1}$ & $/$ & $0.742$ & $0.832$ & $0.595$ & $0.346$ & $0.181$ \\ \hline
$E_{13}^{\ell,2}$ & $/$ & $/$ & $0.051$ & $0.027$ & $0.018$ & $0.013$ \\ \hline\hline
\hline\hline
\end{tabular}
\caption{Overlaps between fermion pairs and first gauge modes for left down quarks. The rows $\omega_{\mu;\ell,0}$ and $\omega^*_{\mu;\ell,1}$ refer to the couplings (mixings taken into account) with these particular modes.\label{tab:quarkcoup}}
\end{table}

We can perform the same procedure for the charged lepton sector, and our
previous conclusions stay more or less valid. In particular, the fact that
$\omega_{\mu;\ell,0}$ don't couple much with $\bar e\mu$ is expected to be
robust, since it depends mainly on the relative equality of all the
$E^{\ell,0}_{nn}$ for low $\ell$.

We now provide the results of exact numerical evaluation at the tree level
for $\Gamma(K_L^0\rightarrow \mu^+ e^-)$ with and without mixings taken
into account. Recall about the chiral suppression of this decay (angular
momentum conservation imposes cancellation of the amplitude for massless
fermions). Thus our result will be of the form $\Gamma\sim
\beta m_{\mu}
^2 m_K R^4 f_K^2$, where $
\beta$ is some dimensionless factor that accounts for the effective
coupling constant which is of order $\sim (g\kappa)^4$. For a SM coupling
$g\sim 10^{-1}$ and an overlap $\kappa\sim 10^{-1} \div 1$ (see Table
\ref{tab:quarkcoup}), we expect $\Gamma\sim R^4 10^{-10}$. This gives a
bound on $R$, but we remind that $R$ plays already a role in the size of
the wavefunctions, so this is only a test \textit{a posteriori} of the
validity of our choice for this parameter\footnote{Nevertheless, if we
consider free $\kappa$'s, we can replace $R$ by $\kappa R$ in
(\ref{ConstR}).}. We could be skeptical about this rough estimation for
$\Gamma$ because we have to sum over all heavy modes (all $\ell$'s), but
remember that (in addition to overlaps reduction) we have a mass
suppression $1/(\ell+1)^4\ell^4\sim \ell^{-8}$ which makes the series
rapidly converging. Indeed, with mixing we have $\Gamma\cdot 10^{10}/R^4=
2.24$, $3.78$, $4.12$, $4.18$ and $4.18$ for $\ell_{\text{max}}=1$, $2$,
$3$, $5$ and $10$ respectively. Without mixing we get, for the same
$\ell_{\text{max}}$, $\Gamma\cdot 10^{10}/R^4= 3.31$, $5.34$, $5.72$,
$5.78$ and $5.78$. It gives the following limits on $R$:
\beq \frac{1}{R}
> 51 (55) \ \text{TeV}
\label{ConstR}
\eeq with and without mixing (for the experimental limit
\cite{branching} on the branching ratio Br$ < 4.7 \times 10^{-22} $),
which
is well below the value $R=100$ TeV assumed in this realisation of the
model.
The model with parameters of Ref.~\cite{Frere:13} (the mass of the new
family-changing vector boson there is $\sim 142$~GeV) is therefore
self-consistent. To obtain a precise lower bound on $R$ for all models,
one needs to perform additional numerical work which is beyond the scope
of the present note. Other rare processes may also be analyzed
\cite{Nemkov}.

\subsection{Collider processes}

Let us briefly comment on the collider phenomenology. At the LHC, our
massive bosons $\omega_{\mu;11}$ could mediate flavour violating processes
if their scale is within the  energy reach of the accelerator --
which would assume an hypothetical geometry where $\kappa  \approx 0.1$.
The typical signature would be a
lepton-antilepton pair with large and opposite
transverse momenta. This is very similar to Drell-Yan pair production for
which a typical feature is the suppression of the cross section with
increasing of the resonance mass at a fixed center-of-mass energy. Note
also that, since we are dealing here with proton-proton collisions, we
expect a dominance of $(e^-\mu^+)$ and $(\mu^- \tau^+$) over $(e^+\mu^-)$
and $(\mu^+\tau^-)$. Indeed the former processes can use valence quarks
($u$ and $d$) in the proton, while the latter involve only partons from
the sea.

A detailed evaluation of the expected number of events at LHC requires
numerical simulation to which we will return in a future note. At this
point however, it already is possible to compute the width of the
$\omega_{\mu;11}$ boson thanks to (\ref{eff4D}). Note that for these
energies, it is more coherent to use $b_{\mu}$ and $\omega_{\mu}^3$
instead of $z_{\mu}$ and $a_{\mu}$. If we neglect possible model-dependent
scalar interactions, we have\footnote{We also neglect masses of all
fermions and therefore mixings are irrelevant.}
\beq \Gamma(b_{\mu;11}
\rightarrow \text{all})=\frac{\sqrt{2}}{R}\frac{g'^2}{32\pi} \left(y_e^2
A_e+ 2 y_L^2 A_L +y_u^2 A_u +y_d^2 A_d+ 2y_Q^2 A_Q\right)\nn
\eeq and \beq \Gamma(\omega^3_{\mu;11}
\rightarrow \text{all})=\frac{\sqrt{2}}{R}\frac{g^2}{64\pi}
\left(A_L+A_Q\right)\nn
\eeq for $A=((E^{1,1}
_{12})^2+(E^{1,1}_{23})^2)$. According to \cite{Frere:04b}, we expect
$\Gamma/\text{GeV}\sim \kappa^2 M/M_Z$, thus $\Gamma\sim 10^{-1}$ TeV. The
exact numerical values for our example are $\Gamma(b_{\mu})=0.44$ TeV and
$\Gamma(\omega^3_{\mu})=0.67$ TeV.

\section{Conclusions and perspectives}\label{sec:Conc}

We have discussed the gauge sector of a successful extra-dimensional model
for masses and mixing of quarks, charged leptons and neutrinos. It is
important for quantitative experimental predictions of the model. Further
details of the warped-geometry case will be discussed elsewhere.

We dedicate this paper to the birthday of Valery Rubakov who is not only
an appreciated pioneer of large extra dimensions. He was a supervisor for
two of us (M.L.\ and S.T.), but he is more than a teacher. He continuously
sets a very high level in his studies and in the works of his school, but
also in his everyday and social life. We are trying to use this level as a
benchmark. Last but not least, it was Valery who initiated the first
contact between J.-M.F., M.L.\ and S.T. in 1999, which resulted in the
development of the branch discussed here.

This work is funded in part by IISN and by Belgian Science Policy (IAP
"Fundamental Interactions").
The work of M.L.\ and S.T.\ (elaboration of the model of the origin and
hierarchy of
masses and mixings in the context of new
experimental data) is supported by the Russian Science Foundation, grant
14-22-00161.

\end{document}